\documentstyle[floats,aps]{revtex}
\input epsf
\begin{document}
\def\noi{\noindent}
\def\BRA{\left\langle}
\def\KET{\right\rangle}
\def\LK{\left(}
\def\RK{\right)}
\def\LBK{\left\lbrack}
\def\RBK{\right\rbrack}
\def\LB{\left\lbrace}
\def\RB{\right\rbrace}
\def\be{\begin{equation}}
\def\ee{\end{equation}}
\def\bea{\begin{eqnarray}}
\def\eea{\end{eqnarray}}
\def\r{{\bf r}}
\def\const{\rm const\ }

\draft
\twocolumn[\hsize\textwidth\columnwidth\hsize\csname
@twocolumnfalse\endcsname

\title{Correlated random band matrices: localization-delocalization transitions}  
\author{Martin Janssen$^{1}$ and Krystian Pracz$^{2}$
\\\ $^1$Institut f\"ur Theoretische Physik III, Ruhr-Universit\"at Bochum,
\\\ 44780 Bochum, Germany
 \\\ $^2$ Institut f\"ur Theoretische Physik, Universit\"at zu K\"oln,
  \\\ Z\"ulpicher Str. 77, 50937 K\"oln, Germany}

 \date{November 29}
\maketitle

\begin{abstract}
We study the statistics of eigenvectors in correlated random band
matrix models. These models are characterized by two parameters, the
band width $B(N)$ of a Hermitian $N\times N$ matrix and the correlation
parameter
$C(N)$ describing correlations of matrix elements along diagonal lines.
The correlated band matrices show a much richer phenomenology than models without correlation
as soon as the correlation parameter scales sufficiently fast with matrix size. In particular,
 for $B(N)\sim \sqrt{N}$ and $C(N)\sim \sqrt{N}$,  the model
shows a localization-delocalization transition of the quantum Hall
type. 
\end{abstract}

 \vskip 2pc]

\section{Introduction}
In the theory of random Hermitian matrices \cite{Guhr98}  two
robust types of statistics are found  in the limit of
infinite matrix size (denoted here as 'thermodynamic limit'). First,
the Wigner-Dyson statistics describing systems that become ergodic  in the thermodynamic limit and have an
incompressible, correlated spectrum and  Gaussian distributed,
uncorrelated amplitudes of the corresponding eigenstates (see
Fig.~1a).  Since we  do not consider  further symmetry constraints we focus on the matrix
ensembles denoted as class A  in the classification of
\cite{AZ96}. A convenient representative of its ergodic limiting ensemble is given by the Gaussian unitary matrix ensemble GUE. 
The second  robust statistics is the  Poisson statistics with eigenstates, localized on
certain basis states (sites), and with an compressible, uncorrelated
spectrum (see Fig.~1b).
\begin{figure}
\epsfysize=6.5cm
\centerline{\epsfbox{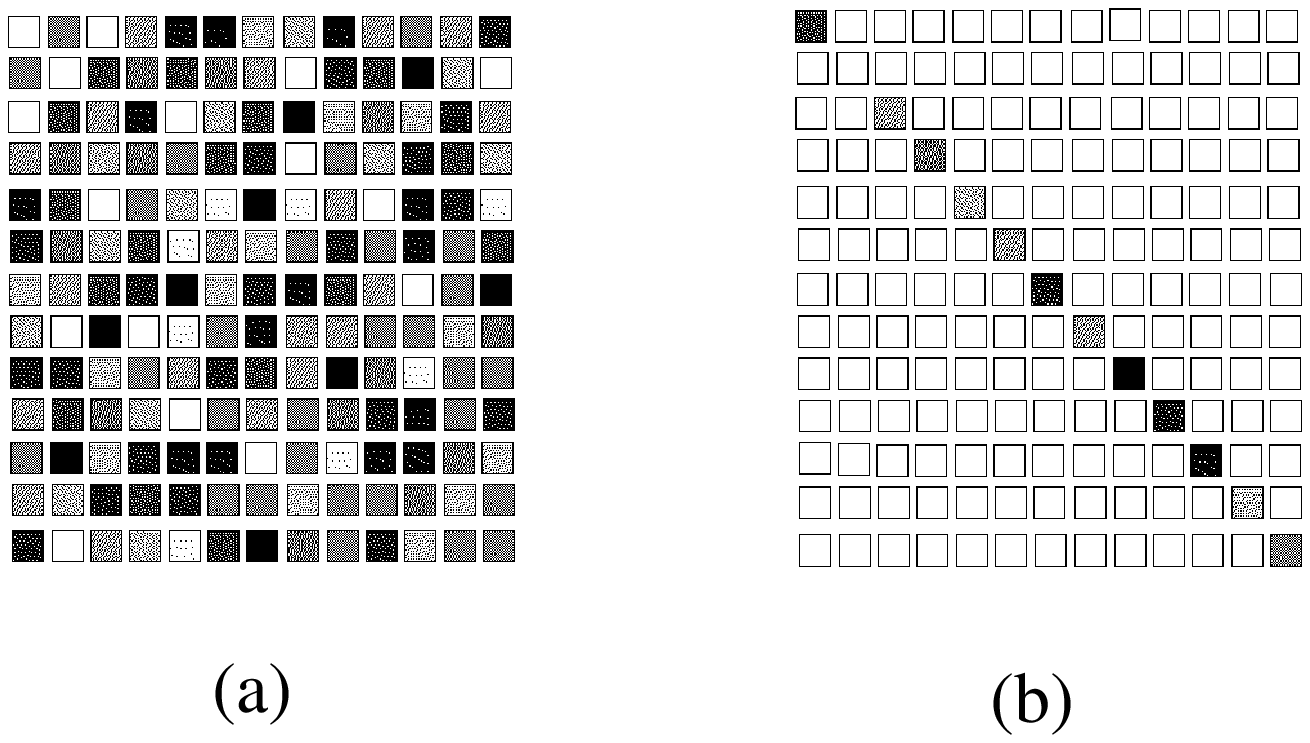}}
\caption{Visualization of  typical random matrices obeying 
Wigner-Dyson  statistics (a),  Poisson statistics (b).  
The intensity  represents  the absolute value of matrix elements.}
\end{figure}
Real complex quantum systems, represented by  random Hermitian matrices, can show a
crossover between Wigner-Dyson and Poisson statistics or, in some
cases, a true quantum phase transition with novel 'critical'
statistics. A well known example is the 3D Anderson model (see  Fig.~2)
\begin{figure}
\epsfysize=6.5cm
\centerline{\epsfbox{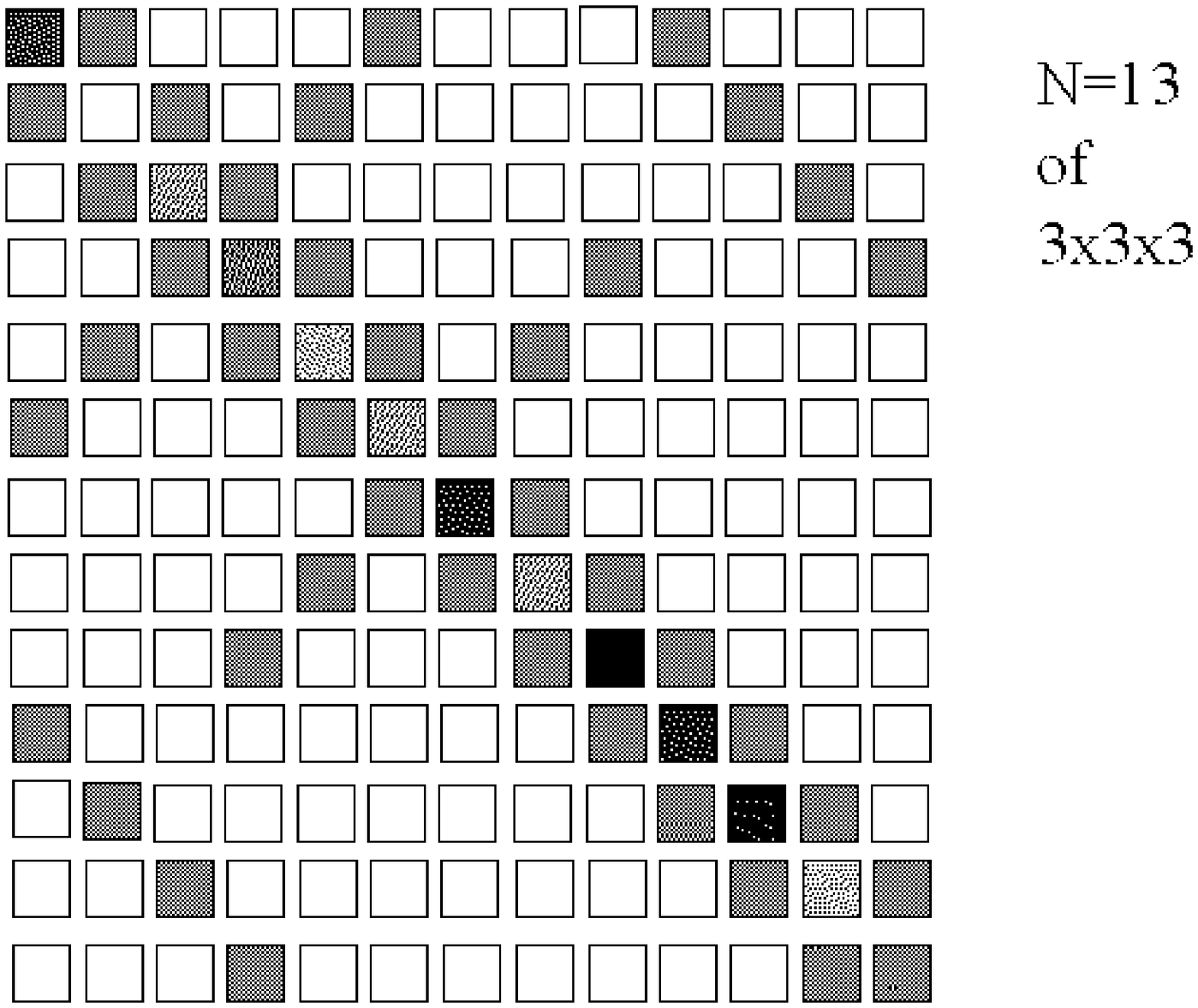}}
\caption{Visualization of a typical random matrix obeying the statistics of a 3D Anderson model.
The intensity  represents  the absolute value of matrix elements.}
\end{figure}
describing the motion of independent electrons on a 3D lattice with
random uncorrelated on-site disorder. Below a certain critical value
of disorder, in the thermodynamic limit, all states at the energy band
center are infinitely extended (delocalized) in space, while for
larger disorder all states are spatially localized.  Instead of
changing the disorder, one can change the energy within the energy
band, keeping the disorder fixed at low values. Again, a transition
from localization (band tails) to delocalization (band center) occurs.
It is worth mentioning, that the 
average density of states (DOS) is non-critical, i.e.
it stays smooth across the localization-delocalization (LD) transition.
Although these statements are substantiated by analytical as well as
numerical work (for reviews see \cite{MK93,J98}), the special structure
of this matrix ensemble (composed of a sparse, but deterministic
matrix and a random diagonal matrix) has prohibited, so far, a
rigorous proof of these statements.  Another well known system with a
transition from localized to critical states is the two-dimensional
(2D) quantum Hall system (for reviews see \cite{H94,JVFH94}) which we will
describe briefly later. Furthermore, several matrix ensembles modeling
the motion of  2D disordered electrons undergoing (time-reversal
symmetric) spin-orbit interactions are known to display a
LD transition (see e.g. \cite{MJH98}).
In all of these realistic matrix ensembles the statistics at
criticality represents an unstable fixed point under increasing system
size (i.e.~matrix dimension), which means that any slight shift away
from the critical value of the energy, say, will drive the system into one
of the stable matrix ensembles, Wigner-Dyson for the
delocalized states and Poisson  for localized states.  The
critical ensembles are characterized by correlated spectra, but with a
finite compressibility. Furthermore, critical eigenstates are
multifractal and the multifractal exponents are related to the
compressibility of the spectrum (for a review see \cite{J98}).

It is desirable to study matrix ensembles with simple construction
rules and to ask for necessary ingredients in order to have a LD
transition.  Also in quantum chaos the interest in crossover
ensembles has grown \cite{C99}.  In that context the Rosenzweig-Porter
model \cite{RP60} was studied as a toy-model for the crossover. It
is defined as a simple superposition of a Poissonian and a
Wigner-Dyson matrix. It has been shown rigorously that, by choosing
the superposition in an appropriate way, novel critical ensembles
emerge, but the spectral compressibility is identical to the Poisson
ensemble and states are not multifractal (see \cite{JVS98} and
references therein).  Another well studied matrix ensemble is that of
random band matrices (RBM) with uncorrelated elements. The band width
$B$ describes the number of diagonals with non-vanishing elements. For
$B\sim N^s$, with $s > 1/2$, one recovers the Wigner-Dyson statistics.  Such band matrix
models have been discussed in the context of the 'quantum kicked
rotor' problem \cite{I90} and have been studied extensively in a
series of papers by Mirlin, Fyodorov and others (for review see
\cite{MF94,M99}). It turned out that, in particular for $B\gg 1$, all
states are localized with a localization length (in index space)
$\xi\sim B^2$. For fixed $B$ one has therefore a crossover from
Wigner-Dyson to Poisson statistics as $N$ is taken from values much
smaller than $B^2$ to values much larger than $B^2$, and $B^2/N$ is
the relevant parameter for a scaling analysis of data.  Superpositions
of such random band matrices with random diagonal matrices have been
studied in the context of the 'two-interacting particle' problem (see
e.g. \cite{Sh94,Fr98}), however these ensembles do not show novel
critical behavior as compared to the Rosenzweig-Porter model.

In fact, only few simply designed matrix ensembles are known to become
critical with multifractal critical states (see \cite{Mir97,Kr98}),
for example 'power law' band matrix ensembles, where the
strength of (uncorrelated) matrix elements falls off in a power law
fashion in the direction perpendicular to the central diagonal. The
critical cases occur for the power law behavior $\sim x^{-1}$ of the
typical absolute values of matrix elements \cite{Mir97,BV}.  It is,
however, important to notice a significant difference to realistic
critical ensembles: there is no LD transition within the spectrum; iff
parameters are fixed to critical values all states are critical.
 
In this paper we study correlated random band matrix (CRBM) ensembles
and, with the assistance of numerical calculations, argue that these
ensembles can lead to a LD transition within the spectrum.  A new
parameter $C(N)\sim N^t$ describing the correlation of certain matrix
elements is introduced for random band matrices. For $B(N)\sim
\sqrt{N}$ states are localized outside of the energy band center and a
LD transition at the energy band center occurs provided $1/2 \leq t
\leq 1$.

A major
motivation for studying these ensemble originates from the theory of
the integer quantum Hall effect \cite{JVFH94}.  The plateau to
plateau transition in the quantum Hall effect can be captured in
models of non-interacting 2D electrons in a strong magnetic field and
a random potential, referred to as quantum Hall system (QHS).  In the
one-band Landau representation the Hamiltonian is represented as a
random matrix with two characteristic features. (i) The matrix elements
decay perpendicular to the main diagonal in a Gaussian way. (ii) No
correlations exist between elements on distinct 'nebendiagonal' lines,
but Gaussian correlations exist along each of the nebendiagonals.
These features led to the introduction of the 'random Landau model'
(RLM)  to study critical properties of
QHSs (see e.g. \cite{H94}). The original purpose  of constructing the RLM was to avoid 
explicite calculations of matrix elements starting from a randomly chosen
disorder potential and to  directly generate the matrix elements  as random numbers that
fulfill the statistical properties  (i) and (ii).  
As will be explained in more details below, the 
corresponding CRBM model simplifies the RLM further, in as much as a
sharp band width is introduced and correlations along nebendiagonals
are idealized and cutoff after a finite length.

Recently, matrix ensembles with correlated matrix elements attracted
some interest \cite{FK99} in the context of the metal-insulator experiments in 2D
\cite{Khv94ff}, for which strong Coulomb interaction is believed to be
a necessary ingredient.  It is very interesting that in \cite{Izr98} 1D
models with correlated disorder potentials could, to a large extent,
be solved analytically and that certain correlated disorder potentials  were shown to cause  LD transitions in 1D.
It is not obvious  how to extend the method of \cite{Izr98} to the case of CRBM models. 
 Usually,
correlations in matrix ensembles lead to serious complications in
analytical attacks. For example, in the field theoretic treatment (see e.g. \cite{EfeB97}) of random matrix ensembles
the absence of long ranged correlations is essential to find appropriate
field degrees of freedom that depend smoothly on a single site variable.  In our CRBM models
correlations are introduced  by constraints (a number of matrix
elements are taken to be identical). This may help to reduce 
complications in constructing a field theoretic approach for CRBM models.

In Sec.~2 we give a detailed definition of  the CRBM  and discuss three alternative
interpretations. The investigation of the LD transition is carried out
in Sec.~3 by a multifractal analysis of states for an ensemble that is
expected to fall into the quantum Hall universality class. Our results are in favor of this expectation.
 The 
analysis is carried over to ensembles where correlations are taken to
extreme limits in  Sec.~4.   In Section~5 we present our conclusions.
\section{Correlated Random Band Matrix Model}
Let the elements of a $N\times N$ Hermitian matrix $H$ be written as 
\bea
	H_{kl} &=& x_{kl}+i   y_{kl} \;\; {\rm
 	for}\;\; l > k \; , \\ 
	H_{kk} &=& \sqrt{2} x_{kk} \, ,
\eea
where all non-vanishing  real numbers $ x_{kl}, y_{kl}$ are taken from the same
distribution ${\cal P}$ with vanishing mean and finite variance
$\sigma^2$. We take  the symmetric and
uniform distribution on $[-1,1]$ ($\sigma^2=1/3$). 
With  $B,C$
 being two  integer numbers, called the 'band width' and
 the 'correlation parameter', respectively,
the correlated band matrix ensembles are defined by the following algorithm (I) -- (III)
and are visualized in Fig.~3.

(I) Begin with the main diagonal of $H$ and draw a random integer number $N_1\leq C$, 
 and the random number $x_{11}$ from the distribution ${\cal P}$. Take the first
 $N_1$ values on the main diagonal equal to $\sqrt{2} x_{11}$. Then, draw
 $x_{N_1+1,N_1+1}$ from the distribution ${\cal P}$ and take  successively $C$ elements
 on the main diagonal  equal to $\sqrt{2}(x_{N_1+1,N_1+1})$.  Now, draw
 $x_{N_1+C+1,N_1+C+1}$ from the distribution ${\cal P}$ 
and take successively  $C$ elements
 on the main diagonal  equal to
 $\sqrt{2}x_{N_1+C+1,N_1+C+1}$. Continue with this procedure until the main
 diagonal is filled up.

(II) Now consider the next 'nebendiagonal'. Its first
random element $H_{12}=x_{12}+iy_{12}$ and  a random integer number
$N_{2}\leq C$ are drawn.  Put the first $N_2$ elements of this nebendiagonal equal to $H_{12}$. 
The next $C$ elements on the same nebendiagonal are taken equal to
the random value $H_{N_2+1,N_2+2}$, and so on.

(III) The procedure terminates after the $B$th
 nebendiagonal ($l-k=B$) is filled up.  All other matrix elements ($l-k > B$) are
set to zero. Finally, Hermiticity is installed by taking $H_{l>k}=H^*_{k<l}$.
\begin{figure}
\epsfysize=7cm
\centerline{\epsfbox{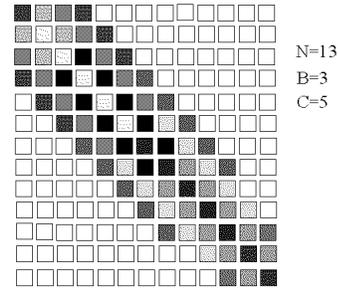}}
\caption{Visualization of  a typical matrix of the correlated random band matrix model. 
The intensity represents  the absolute 
value of matrix elements.}
\end{figure}

For $C=1$ the usual band matrix models (band width $B$) of uncorrelated matrix elements 
are recovered. $C$ and $B$ form the relevant parameters of the CRBM, while
the  value of $\sigma$ is not significant -- it just defines the energy units.
For finite $C>1$  the correlation along (neben)diagonals is a triangular function
of range $C$ and half-width $C/2$. Thus, $C/2$ is a typical distance over which
elements are correlated along (neben)diagonals. 
The spectrum is always distributed in a symmetric way around the center $E=0$
as a consequence of the symmetry of the  distribution ${\cal P}$. 

In the following, we are going to discuss three possible physical interpretations.
The most obvious interpretation relies on the  site representation $\left.\mid l \KET =
(0,0,\ldots,l=1,0,\ldots,0)$. In this representation,  
$H$ describes hopping of particles on a 1D chain  of length $L=N$ (lattice spacing $=1$)
with a maximum distance of hopping equal to $B$. The average hopping probability in 
one instant of time $t$ is ($\hbar=1$, $\BRA \ldots \KET$ denotes the ensemble average)
\be
	\BRA|\BRA k(t)|l(t+1)\KET|^2\KET = \sigma^2 \;\; {\rm for}\;\;
	0\leq
	|l-k|\leq B \, .
\ee
\begin{figure}
\epsfxsize=6cm
\centerline{\epsfbox{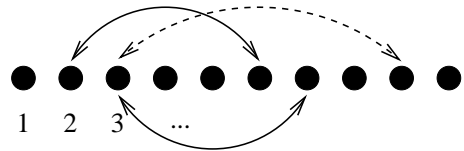}}
\vspace{0.3cm}
\caption{One-dimensional interpretation of correlated band matrices.
Hopping events for distinct hopping distances are uncorrelated.
Hopping events for the same hopping distance are correlated over typically
$C/2$ nearest neighbors. In this example $ B \geq 6$,  $C \geq  3$.}
\end{figure}
The correlation between matrix elements means that  two hopping amplitudes are 
identical, if  the hopping distance is equal, and provided the hopping starts at
sites the distance between which is less than  typically
 $C/2$ (see Fig.~4).

An alternative interpretation results, when the $N$ sites are arranged in a
quasi-1D  (Q1D)  geometry with  $N_c=B+1$ parallel
'channels' in a 'wire' of length $L'=N/(B+1)$.  The lattice
spacing {\it along } the wire, $\hbar$, and the unit of time are taken to be $1$.
Transitions in one unit of time are possible between channels (coupling) and 
along the wire  (hopping). Sites are labeled (see Fig.~5) such that hopping is possible at most  over
one lattice spacing along the wire. 
 Again, hopping (coupling)   is correlated
over typically $\sim C/2$ states, provided the difference between labels is identical.
\begin{figure}
\epsfxsize=5cm
\centerline{\epsfbox{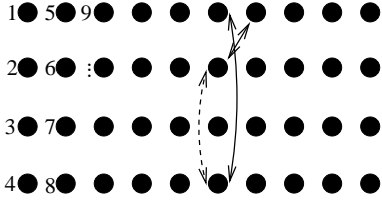}}
\vspace{0.3cm}
\caption{Quasi-one-dimensional interpretation of correlated random band matrices.
Sites are labeled such that for a fixed cross section of the wire all channels are filled up
with sites, before one proceeds to the next cross section. 
Hopping events along the wire and couplings within each cross section are 
uncorrelated, if  label distances are distinct. For identical label distances
they  are correlated over typically
$C/2$ sites. In this example $B=3$,  $C\geq 2$.}
\end{figure}
A less obvious interpretation of CRBM models arises when
quantum Hall systems are studied  in a one band  Landau representation.
A quantum Hall system is described by a one-particle Hamiltonian
of 2D electrons (charge $-e$, mass $m$) 
in the presence of a strong perpendicular magnetic  field ${\cal B}$
and a random potential $V(x,y)$. In Landau gauge the Hamiltonian  reads
\be 
	H=\frac{1}{2m}\LK p_x^2 + (p_y + e{\cal B}x)^2\RK + V(x,y) \, ,
\ee
where $(p_x,p_y)$ is the canonical momentum with respect to the Cartesian coordinates $(x,y)$.
For periodic boundary conditions in $y$-direction (length $L_y$) the kinetic energy
forms a highly degenerate harmonic oscillator
problem ($p_y$ is conserved) that is diagonalized by Landau states \cite{Lan29} 
$\left.\mid n,l\KET$. Here $n=0,1,2,3,\ldots$
labels the Landau energies $E_n=\hbar \omega_c (n+1/2)$ ($\omega_c=e{\cal B}/m$ cyclotron energy),
and $l=0,1,2, \ldots $ labels center coordinates of the degenerate Landau states. 
The Landau states are separated into plane waves in $y$-direction with quantized momentum $q_l$ and into
oscillator wave functions centered at $X_l=-\lambda^2 q_l$, where 
$\lambda=\sqrt{h/(e{\cal B})}$ is the characteristic 'magnetic length'. As long as the typical
 values of the random potential are 
much smaller than the cyclotron energy, one can study the full eigenvalue problem of the low
lying 'Landau bands' approximately by restricting the Hilbert space to separate Landau levels $n$. 
In particular, for the lowest Landau level the Landau states read  
\be
	\BRA x,y| l\KET = \frac{1}{\pi^{1/4}\sqrt{\lambda L_y}} \exp\LB
	{-i\frac{X_l 	y}{\lambda^2}}\RB\exp\LB{-\frac{(x-X_l)^2}{2\lambda^2}}\RB \, . \label{2.1}
\ee
A convenient recipe to study finite systems of length $L$ in $x$-direction is to use
 'Landau counting' of states, that is to take only those Landau states into account for which the center coordinates $X_l$
fall into the interval $[0,L]$. The total number of states, for an aspect ratio $a=L_y/L$, is 
\be
	N=(a/2\pi) (L/\lambda)^2 \, .
\ee
By shifting the lowest Landau energy to zero, the eigenvalue problem
is defined by the matrix
$H_{kl}=\BRA k |V| l\KET$ which reads
\bea
	H_{kl}=\frac{e^{-\frac{1}{4\lambda^2}(\Delta X)^2}}{\sqrt{\pi}\lambda}
	\int\limits_{-\infty}^{\infty} dx \, \tilde{V}(x; \Delta X)
	e^{-\frac{1}{\lambda^2}(x-X)^2}\, ,\\
	\tilde{V}(x;\Delta X)\equiv
	L^{-1}_y\int\limits_{-{L_y/2}}^{L_y/2}  dy\, V(x,y) e^{i y\Delta X/\lambda^2 }\, ,\label{HKL}
\eea
where $X=(X_k+X_l)/2$ and $\Delta X= X_k-X_l$. 
These matrix elements  form a random $N\times N$ matrix, the  elements of which are 
composed by a Fourier transformation of the random potential in $y$-direction
and a Gaussian weighted averaging of the potential over a magnetic length in $x$-direction.
Crucial for the structure of the matrix is
the fact that, within the distance of one  magnetic length $\lambda$, a number
of $N_\lambda \sim N(\lambda/L)$ different Landau states can be situated (see Fig.~6). Thus, for a
 constant aspect ratio, this number increases as $N_\lambda \sim \sqrt{N}$ in the thermodynamic limit.
\begin{figure}
\epsfxsize=5cm
\centerline{\epsfbox{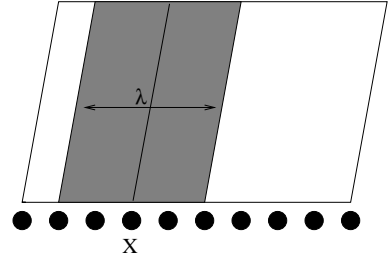}}
\vspace{0.3cm}
\caption{Interpretation of correlated band matrices as quantum Hall system.
Bare states are interpreted as Landau states on a 2D plane. The qualitative properties of Landau
 states are shown: plane waves in $y$ direction situated at center coordinate $X_l$, extended in 
$x$-direction over approximately one magnetic length $\lambda$ corresponding to
$N_\lambda$ neighboring states.}
\end{figure}
This  leads to correlations between matrix elements along nebendiagonals over 
a range of $N_\lambda$ states. This range increases when the  disorder potential is 
correlated in real space over distances exceeding the magnetic length.
The correlation between matrix elements on distinct (neben)diagonals is negligible, since
the correlator contains the superposition of $\sim N_\lambda$ random  phase factors. 
The matrix elements decay perpendicular to
 the main diagonal due to the Gaussian decay of Landau states. For disorder potentials with
a spatial correlation length $d\gg \lambda$  the decay sets in earlier, because
the Landau states are orthogonal. A quantitative analysis shows (see \cite{H94}) 
that  the Gaussian band-width is 
\be
	\tilde{B}= \frac{1}{\beta}\sqrt{\frac{aN}{\pi}}\label{bandwidth}
\ee
and the Gaussian correlation length along (neben)diagonals is
\be
	\tilde{C}= \beta \sqrt{\frac{aN}{\pi}} \, .\label{corrlenght}
\ee 
Here $\beta \geq 1$ is a parameter that is controlled by the potential correlation length $d$,
\be
	\beta\equiv \frac{\sqrt{d^2 +\lambda^2}}{\lambda}\, .\label{beta}
\ee
For finite and fixed values $d,a$, in the thermodynamic limit,
$\tilde{B}(N)$ and $\tilde{C}(N)$ increase as $\sqrt{N}$ to infinity. A matrix  for 
QHSs   with the statistical properties described above is denoted as 'random Landau matrix' (RLM) (see \cite{H94}). 

Our CRBM simulates the  RLM in as much, as it has the same qualitative
features of a  finite band width $B$ and 
a  correlation length, being typically $C/2$. The quantitative differences are  that, in the CRBM, the matrices  have a sharp
band width $B$, instead of a Gaussian band width $\tilde{B}$, and that the correlation function is a triangle with a half width $C/2$,
instead of a Gaussian with a half width $\tilde{C}$. In the thermodynamic limit, however,
we expect that these differences should be insignificant for the statistics of eigenvalues and
eigenvectors, provided the parameters $B(N)$ and $C(N)$ scale in the same way with $N$ as
the parameters $\tilde{B}(N)$ and $\tilde{C}(N)$, respectively.
Note, however, that in the RLM the ratio $\tilde{C}(N)/\tilde{B}(N)= \beta^2 $ is bounded from below by 1, 
while in the CRBM model we are free to choose any value for $B(N), C(N)$.
\section{Multifractality and Spatial Correlations}
To study the LD properties of matrix ensembles
numerically, one can follow a number of different strategies. The most
efficient is to  analyze only the eigenvalue statistics.
Although the eigenvalues encode most of the relevant information about
LD properties,  the statistics of wavefunction amplitudes
is more direct.
Localization, delocalization  and even criticality of states can be qualitatively
distinguished already by  inspection of plots of the squared amplitudes of wavefunctions (see e.g. Fig.~7).
Critical states $\psi(\r)$ are characterized by having a  multifractal 
distribution of its squared amplitudes ${\rm prob}(\r)\equiv |\psi(\r)|^2$. 
 This spectrum becomes independent of system size and
is universal for all of the critical states in the thermodynamic limit
(for a review see \cite{JanR94}) or, more generally, follows a universal
distribution ( see \cite{Par99}). In particular, the geometric mean is a convenient measure
of a typical probability and scales as 
\be
      {\rm prob}_{\rm typ} = \exp \BRA \ln({\rm prob(\r)})\KET \sim L^{-{  \alpha_0}}\, ,
\ee
where the deviation $\alpha_0-d\geq 0$ of the fractal dimension $\alpha_0$ from
the Euclidean dimension $d$ signals multifractality and is the most sensitive critical exponent
of the LD transition. Although a critical state is extended all over the system, 
it fluctuates strongly and has large regions
of low probability which results in the stronger decay of typical amplitudes as compared 
to homogeneously extended states.
A  quantity that is closely related to $\alpha_0$ is the exponent $\eta$ of long ranged spatial correlations
$\BRA  {\rm prob}(r) {\rm prob}(0)\KET \sim r^{-\eta}$ \cite{Weg80,ChaDa88}. It 
fulfills scaling relations to the fractal dimension of the second moment of ${\rm prob}(\r)$ \cite{PoJa},
and also to the compressibility of the eigenvalue spectrum \cite{KraLerCha} (see also \cite{Poly}). 
As a crude estimate   (based on a log-normal approximation\cite{PoJa} to the 
distribution of ${\rm prob}$) one has
$\eta\approx 2(\alpha_0-d)$. 

In this work we focus on wavefunction statistics and the determination of $\alpha_0$.
One should, however,  be careful when drawing  conclusions 
from the calculation of fractal exponents for finite matrix sizes. Such calculations should be assisted
 by inspection of states,  and  
one should study the dependence on the matrix size $N$.
For example, states with localization lengths that are small, but not very small 
compared to system size tend to produce larger values of $\alpha_0$ because  parts 
of the wave function have low amplitudes. 
This can be seen easily in plots of the corresponding squared amplitudes.
From the linear regression procedure that allows to determine $\alpha_0$ one cannot distinguish such  behavior 
from true multifractality, as long as  the system sizes
cannot be made much larger than the localization length.  Furthermore, one should distinguish
between spatial, energy, and ensemble averages. In practice, we first perform the
spatial average for a fixed wavefunction to determine the exponent  $\alpha_0$
for a finite matrix size $N$ by the box-counting method (see e.g. \cite{PoJa}).
This can be done for many states in the critical region (which has finite width for finite $N$)
and we average over these states. Finally, an average over different realizations (ensemble average) 
can be performed. It is not obvious, that the order of these different procedures commutes 
since  the exponent fluctuates from state to state.
The question about the self-averaging of the fractal exponents was recently addressed in \cite{Par99}
 and it was claimed that exponents follow a universal scale independent distribution function 
in the thermodynamic limit, rather than  being self-averaging. 
We do not investigate this question here. For our finite systems the exponents are fluctuating
anyway and we consider averages (over typically 100 states) as explained above.  

For a number of different models of  QHSs the exponent $\alpha_0$ has been determined, e.g 
$\alpha_0=2.28 \pm 0.02$ in \cite{Pra96}. Other authors (e.g. \cite{ALFA0})
find values between $2.27\pm 0.1$ and  $\alpha_0=2.29 \pm 0.02$. The largest system sizes 
studied were about $L\approx 200 \lambda$ leading
to matrix dimensions $N$ of about $10^4$.

For the sake of a direct comparison of wavefunction plots between our CRBM models and the RLM
we took the 2D Landau representation of Eq.~(\ref{2.1}) with a Landau counting
for an aspect ratio $a=1$. The eigenvalues and eigenstates
were calculated numerically exploiting the band structure of the matrices. 
To be as close as possible to a real QHS with an aspect ratio $a\approx 1$ and a short ranged
random potential, we took $B(N)=\tilde{B}(N,a=1,\beta=1)= \sqrt{N/\pi}$ 
and $C(N)=(1/2)\tilde{C}(N,a=1,\beta=1)=(1/2)\sqrt{N/\pi}$ 
(e.g. the largest matrices had parameters
$N=6400$, $B=45$ and $C=90$). We refer to this choice of parameters as the ''standard quantum Hall case''

Our findings for
the standard  quantum Hall case of the CRBM model can be summarized as follows.
Almost all states are localized (see Fig.~7a). Only those around  the
energy band center $E=0$ are  multifractaly extended (see Fig.~7b).
\begin{figure}
\epsfysize=13cm
\centerline{\epsfbox{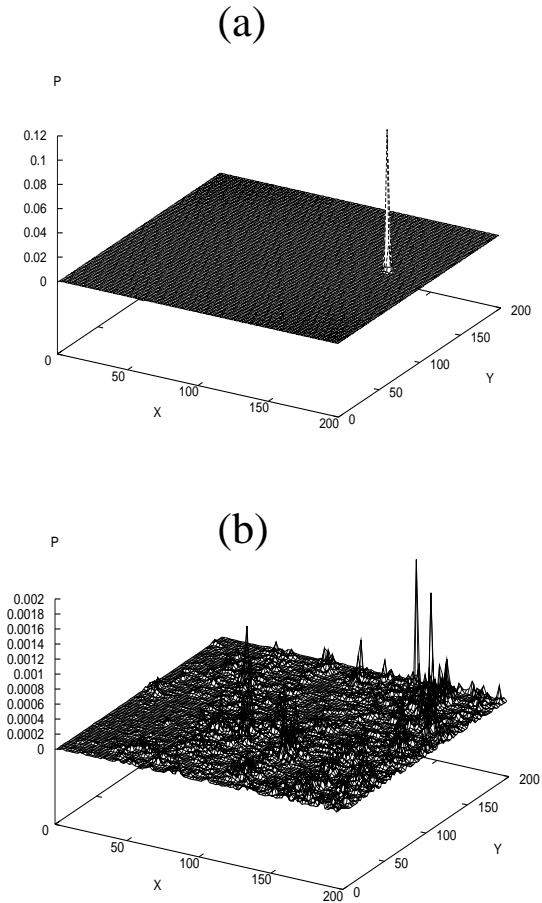}}
\vspace{0.3cm}
\caption{Squared amplitudes of a localized state (a), and a critical state (b),  in the correlated random band matrix 
 model for the  standard quantum Hall case. States are represented  in two-dimensional Landau representation.}
\end{figure}
For finite $N$ there is a small energy band of extended states with
localization length $\xi \geq L$. The energy band width of extended states, $\Delta_c$, shrinks with increasing
$N$ with a critical exponent related to the divergence of the localization length at the band center. 
To determine this  critical exponent it would be sufficient to determine $\Delta_c$ as a function of system size.
 However, $\Delta_c$ can be defined only by considering an 
appropriate scaling variable, e.g the participation ratio, that increases above some threshold value 
when the states become extended. In finite systems scaling variables are strongly fluctuating and 
one has to consider distribution functions and/or  appropriate 
typical values. We did not try to calculate this exponent precisely, but 
only convinced ourself that the typical numbers of clearly extended states was comparable
 to those of realistic quantum Hall systems with $\tilde{B}(N)=B(N)$, $\tilde{C}(N)=2C(N)$.

The fluctuation of $\alpha_0$ determined by the box-counting for individual states can bee seen 
in Fig.~10. The average over 100 different extended states of a $N=6400$ standard quantum Hall 
case  yields $\alpha_0=2.26\pm 0.02$. This value 
  is close to the value $\alpha_0=2.28\pm 0.02$ obtained by the same  averaging procedure
for an original quantum Hall system in \cite{Pra96}. We therefore conclude, that the CRBM
shows indeed a LD transition reminiscent of quantum Hall systems. It seems likely  that, in the thermodynamic
limit, the critical behavior of CRBMs in the standard quantum Hall class is actually identical to that of the 
quantum Hall universality class, because 
the essential $N$ dependence of the relevant parameters, band width $B(N)\sim \sqrt{N}$ and 
correlation parameter $C(N)\sim \sqrt{N}$, are identical to those of realistic QHSs in Landau representation. 

So far the comparison of multifractal exponents was based on the wave function statistics without any reference to spatial
correlations. Therefore, a more ambitious comparison between the CRBM and a QHS concerns the critical 
exponents of spatial correlations and their scaling relations to the multifractal exponents. 
In a multifractal state the $q$-correlator $\BRA  {\rm prob}^q(r) {\rm prob}^q(0)\KET \sim r^{-x(q)}$ has fractal dimension 
\be
	x(q)=2\Delta(q)-\Delta(2q) \label{scalrel}
\ee
where $\Delta(q)$ are the usual fractal exponents of the 
$q$-moments $\BRA {\rm prob}^q(\r)\KET \sim L^{-\Delta(q)}$ (for review see \cite{J98}).

We find for the spatial correlations of the standard quantum Hall case the scaling exponents shown  in Fig.~8.
They are compared with the data that follow from the spectrum $\Delta(q)$ and Eq.~(\ref{scalrel}). The spectrum
$\Delta(q)$ was calculated by the box counting-method. 
\begin{figure}
\epsfysize=8cm
\centerline{\epsfbox{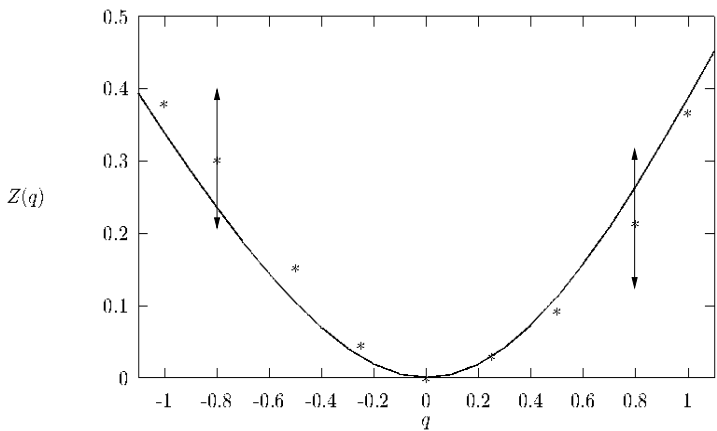}}
\caption{Numerically obtained correlation exponents $z(q)$ (symbols) in comparison with the data following from
the scaling relation (\ref{scalrel}) (line). Error bars are shown for two representative values.}
\end{figure}
They satisfy, within the numerical uncertainties, the 
scaling relation Eq.~(\ref{scalrel}). Furthermore, the exponents are close to their values for critical states in realistic  
quantum Hall systems \cite{Pra96}, e.g. the exponent of  'anomalous diffusion'  \cite{ChaDa88} is
$x(1)=0.4\pm 0.1$. 

We summarize  this section by stating  that  our multifractal analysis gives  a number of clear indications
that  correlated random band matrices, in the   ''standard quantum Hall''  case, are true  representatives 
of the quantum Hall universality class.

We close this section with an
observation not directly relevant for the questions addressed in this paper, but which may be relevant for those
readers  that  like  to perform  numerical calculations 
with the CRBM model. In realistic QHSs with an aspect ratio $a\not=1$
one observes that  a number of the multifractal states tend to have an orientation along the  direction of smaller width.  
This orientation effect has no influence on the asymptotic statistical properties of the wave function amplitudes as
 $N$ scales to $\infty$. Only corrections to scaling
due to finite system sizes can be different for different aspect ratios. However,
an aspect ratio $a\not=1$ will  influence scaling variables like the  Thouless sensitivity 
$g_{\rm Th}=\delta \varepsilon/\Delta$. It  measures  the change in energy $\delta \varepsilon$ due to a
 change from periodic to antiperiodic
boundary conditions in a given direction, relative to the mean level spacing $\Delta$. It is exponentially 
small for localized states and
typically of order 1 for critically extended states. We calculated this quantity for a realistic quantum Hall
system. It has strong mesoscopic fluctuations (variance $\sim$ mean) and found that the unique maximum 
of its typical values at the
band center, for $a=1$, splits up into two maxima,  for $a\not=1$, symmetric around the band center.  
This behavior is  related to the fact that those wave functions that start to be extended in the direction of 
smaller width are more sensitive to changes in the boundary condition than those that have already 
huge localization lengths and are  uniformly extended in both directions. In the CRBM model 
we observe a similar phenomenon. For the standard quantum Hall case
we actually found a tendency for an orientation into the $y$-direction (with 
 Landau states as defined in Eq.~(\ref{2.1}) and Landau counting for $a=1$)). 
This behavior  changes to an orientation in the  opposite direction under   increasing $C(N)$  by a factor of ${\cal O}(1)$, 
keeping $B(N)$ fixed. In contrast to a realistic QHS
where we can calculate $\tilde{B}, \tilde{C}$ for a truly symmetric situation, $a=1$,
in the CRBM model we do not  know a priori if the choice $C/2=B$ is  appropriate to simulate
a truly symmetric situation, $a=1$. Because the identification of the half-width
value of a triangular correlation function with the half-width of a Gaussian is not strict, taking
a  factor of order unity between them is equally well justified.
The same ambiguity is present in the identification of the band width $B$ with $\tilde{B}$.
Any  change in $B$, $C$ by a factor of order unity can therefore lead to the
orientation effect. As in realistic quantum Hall systems we also found the  splitting of the maximum  in Thouless sensitivities
in the standard quantum Hall case.
This splitting effect may  be  further investigated. 
\section{Tuning the Correlation Parameter $C(N)$}
When the correlation parameter is fixed to a constant $C(N)\equiv C_0$ the CRBM will 
behave like an ordinary random band matrix
when $N\gg C_0$. This case is very well understood (see \cite{MF94,M99}) and 
one knows that a crossover from localization
to Wigner-Dyson delocalization takes place as the band width $B$ is varied.
 The localization length in 1D interpretation is $\xi_{\rm 1D}=
 c B^2$ ($c$ a constant of order unity). Except for the far energy band tails, where states are stronger localized, 
the localization behavior is almost uniform over
the energy band. It is worth noting  that the amplitudes of wavefunctions within the 
central region (where the amplitudes are not
exponentially small) are strongly fluctuating, but they
are {\it not} multifractal in the limit of  large $B(N)$.  The entire distribution of amplitudes is,
 asymptotically in $B$, 
fixed by the value of the ratio $g=\xi_{\rm 1D}/N=c B^2(N)/N$. This ratio is the relevant scaling 
parameter and is denoted 
as ''conductance''. As we have seen in the previous section, the behavior of CRBM changes drastically
when the correlation parameter  increases sufficiently fast with $N$. In the standard quantum Hall case, 
a LD transition takes place within the energy band. To get more insight into the role of the correlation 
parameter we therefore  studied two extreme cases:   
(A)  $C(N)=1$, 
and     (B) $C(N)=N$. In both cases we kept $B(N)\sim \sqrt{N}$ as in the standard quantum Hall case.

 Situation (A) corresponds to the usual uncorrelated random band matrix models 
 with a large localization length of
$\xi_{\rm 1D}\sim B^2 \sim L$ in 1D-inter\-pre\-tation ($\xi_{\rm
Q1D}\sim B \sim L'$ in Q1D-inter\-pre\-tation) and a constant {
'conductance' of order $1$}, $g=\xi_{\rm 1D}/L=\xi_{\rm Q1D}/L'={\rm
const}$.  This model has {\em
no} interpretation as a QHS, since the ratio $C(N)/B(N)\sim N^{-1/2} \ll 1$. 
For better comparison we used the same Landau representation as before and performed a multifractal analysis
of the extended states by the same box-counting method as in the standard quantum Hall case.

Our findings in situation (A) can be summarized as follows.  All states behave similar within the band (except for those in the far tails)
The states are not uniformly extended, but
are confined to strips with a width of  about half the system size
 (see Fig.~9). Within that strip the states are extended and
 they fluctuate  strongly 
in a 'grassy' way. They do not show the self-similar regions   of low amplitude like
 typical multifractal states. 
\begin{figure}
\epsfysize=6cm
\centerline{\epsfbox{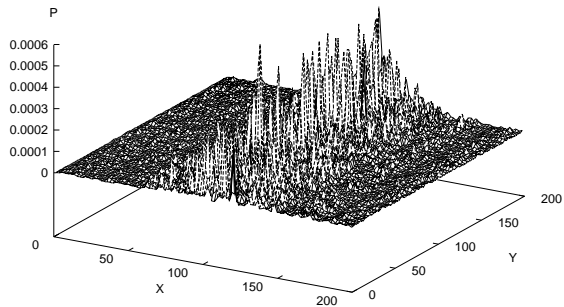}}
\caption{Squared amplitudes of a typical state  in the random band matrix 
 model for situation (A) with vanishing correlation between matrix elements. The  state is  represented  in the two-dimensional Landau representation.}
\end{figure}
This behavior is compatible with the 1D
 (or quasi-1D) interpretation of the uncorrelated random band matrix with
 a localization length  of the order of $L$ ($L'$) and a
 conductance  of order unity. 
We calculated, for $N=6400, B=45$, the exponent 
$\alpha_0^{[2D/C=1]} 
\approx 2.14$ (see  Fig.~10). 
This value must  be taken with care,
 as the states were not extended over the full system. They 
are  localized to an area of about half the system size. Thus, the
 regions of exponentially small amplitudes outside the localization
 center lead to values $\alpha_0> 2$. Taking amplitudes from only the
 localization center  reduces the average of $\alpha_0$, but  fluctuations
 from state to state are strong. We therefore expect that
 $\alpha_0$, measured in the localization center, will  slowly converge to
 $\alpha_0=2$ 
as $B(N)$ increases further with $N$. 
\begin{figure}
\epsfysize=6cm
\centerline{\epsfbox{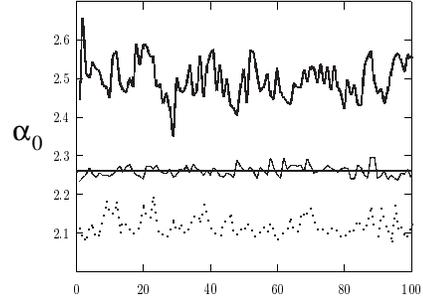}}
\caption{Multifractal exponents for 100 extended states of correlated random band matrix models
in the standard quantum Hall case (middle), in situation (A) (bottom) and in situation (B) (top). The straight line
corresponds to the average in the standard quantum Hall case.}
\end{figure}

Situation (B) deviates strongly  from the usual  uncorrelated random
band matrix models. The elements on each (neben)diagonal are
constant, but uncorrelated for distinct (neben)diagonals. We denote
them by $h_m$ where m=0 labels the main diagonal and positive
(negative) $m$ label the upper right (lower left) lying
nebendiagonals. 

Had  we  set $N=\infty$ in the first place, we could
solve the eigenvalue problem by Fourier 
transformation. Each hopping event over a fixed distance would be translational invariant. Thus, the eigenstates, for $N=\infty$, are
 plane waves 
$\psi_q(l)=e^{iql}$ where $q$ is a quantum wave number that can take any real value. The corresponding eigenvalue is
\be
	E_q=\sum_{m=-\infty}^{\infty} h_m e^{iqm}=h_0 + 2\Re \sum_{m=1}^B h_m e^{iqm} \, .\label{5.5}
\ee
In Landau  representation the plane waves $\psi_q(l)$ transform into wave functions $\psi_q(x,y)$ that are plane waves
in $x$ direction, centered at a center coordinate $Y_q=-\lambda q$, and  have a width of a magnetic length in $y$ direction.

For any finite $N$, however, such solution is not possible, unless  periodic boundary conditions
are implemented in the site representation. To implement them
into our band matrix models  we have to add
 $\sim B^2$ matrix elements in the upper right (and
lower left) corners of the matrix. This would   violate the band structure. We see that, 
 for any  finite $N$, the correlated  band-matrix brakes the translational invariance of hopping events,
and it is not obvious that the states restore this
symmetry when  $N$ goes to infinity. Actually, our finite $N$ results
indicate that the states  will not be plane waves in the center of the
band (see Fig.~11).  Furthermore, a simple perturbative treatment shows that the omission of the $\sim B^2$ elements
in the corners cannot be neglected in the limit $N\to \infty$.

The  CRBM in
situation (B) does also allow for an  interpretation as a quantum Hall system, since
$C(N)/B(N)\sim \sqrt{N} > 1$.  As follows from equations
(\ref{bandwidth} -- \ref{beta}) the potential correlation length $d \sim N^{1/4}$ and
the aspect ratio is large, $a\sim \sqrt{N}$. This translates to the
scaling with system size $L$ as 
\be 
	d\sim L\sim N^{1/4} ,\; L_y=aL
	\sim L^3 \, .  
\ee 
The CRBM in situation (B), thus represents a long quantum Hall strip where
$L_y/L \sim (L/\lambda)^2$ and the random potential can be thought of as being  smooth
over a distance of the width $L$. With periodic boundary conditions
in $x$-direction one would again conclude that eigenstates are plane waves
in $x$-direction, labeled by $N$ different center coordinates $Y_q$ ($q$ is an integer times $2\pi/L$) in
$y$-direction, and the eigenvalues $E_q$ would be  determined by the
value of the random potential at center coordinate $Y_q$. This scenario  is also 
consistent with Eq.~(\ref{5.5}), because the Fourier transform of the
random potential at $V(Y_q)$ yields the matrix elements $h_m$ (see also Eq.~\ref{HKL}). In the absence
of periodic boundary conditions the situation changes.
For energies far from the band center one expects, that the corresponding eigenstates 
are localized on equipotential contours  of the random potential and are centered at some value 
$Y_q$. However, close to the energy band center
eigenstates become extended  and one typical eigenstate
is shown in Fig.~11.
\begin{figure}
\epsfysize=6cm
\centerline{\epsfbox{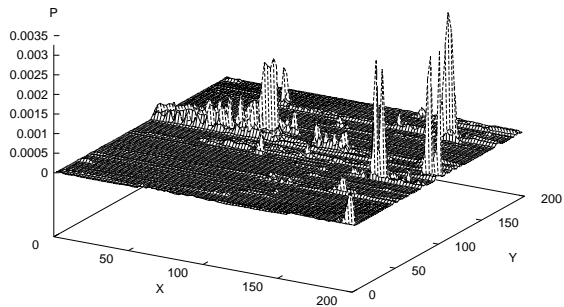}}
\caption{Squared amplitudes of a typical state  in the random band matrix 
 model for situation (B) with strongest correlation between matrix elements. 
The  state is  represented  in the two-dimensional Landau representation.}
\end{figure}
 Although this state has a preferred orientation in $x$-direction it is by no means  localized to a
 small region in  $y$-direction. It fluctuates strongly, it has non-vanishing values all over the system, and it also
 shows  large areas of low probability. Therefore,
the multifractal exponent $\alpha_0$ is larger than in the standard quantum Hall situation.

Let us try to give heuristic arguments of how to estimate the value of $\alpha_0$.
 For that purpose we recall that,
   quite generally  $\xi_{\rm Q1D}$ 
is of the order of the number
of transverse modes 
$N_c$ times the relevant scattering length $l$ (for a discussion see e.g. \cite{J98}). 
In our situation $N_c=N$ and $l\approx d\sim L$.
Therefore, the quasi-1D localization length is estimated to be $\sim  L_y^{5/3}$, and it 
 is much larger than $L_y$ \cite{Note}.
We may thus assume that the state is critical and has, in the strip-representation, a value
 $\alpha_0\approx 2.26$ when the fractal analysis
is restricted to sizes much larger than $d\sim L$. Recall that  we have chosen  the Landau representation 
corresponding to an aspect ratio
$a=1$. Therefore,  the value of $\alpha_0$  found by box counting in that representation must be  different. 
The box counting method  uses squares of size $l^2$ in the 2D Landau representation with $a=1$. This 
corresponds to
{\it rectangular boxes} in the strip-representation, where the length in $y$ direction scales as the
 third power of the length in
$x$-direction. Thus, the ''effective volume'' is  $l^4$.
By  this reasoning  $\alpha_0 (C=N) -2$ will be, in the $a=1$ representation, two times larger than in the
 strip-representation, and
we may expect that  we should find 
$\alpha_0^{[C=N]}\approx 2.54$ by box-counting in 2D Landau representation with $a=1$. Indeed, 
this estimate is compatible   with our findings as displayed in Fig.~10.
\section{Conclusions}
We have studied a novel type of matrix models, the correlated random band matrices. 
We used numerical diagonalization and performed
a multifractal analysis to analyze the localization-delocalization properties of such matrix models
 in the thermodynamic limit of infinite matrix size.

The  parameters of correlated band matrices are the band width $B$ and the correlation parameter $C$.
We offered three interpretations: (i) independent quantum particles on  a  one dimensional chain with correlated hopping,
(ii) independent quantum particles on a quasi-one-dimensional
strip  with correlated coupling of channels, and (iii) in some range of its parameters
the models 
resemble two dimensional quantum Hall systems.  
  For $B\sim C\sim \sqrt{N}$ a transition from localized to  critical states in the band center occurs, and the corresponding
 critical exponents are  close to those of real quantum Hall systems.
Furthermore, 
  we found the following qualitative behavior when keeping the band
width $\sim \sqrt{N}$ constant: A reduction of correlations suppresses
multifractality (i.e. criticality) at the band center and finally, for $C=1$, the
ordinary non-critical random band matrix ensemble is reached which shows localization lengths $\xi\sim B^2$. 
  Increasing correlations beyond $C\sim \sqrt{N}$, the transition to
critical states in the band center remains, however their
multifractality seems to be more pronounced. 
The fractal critical exponent for extreme correlations, $C(N)=N$,  turned out to be compatible with a  heuristic estimate.

Therefore, our numerical results suggest that the correlated band matrix models
show transitions from localization to critical delocalization on approaching the energy band center, provided
the band width scales like $B(N)\sim \sqrt{N}$ and the correlation parameter scales like $C(N)\sim N^t$ with $1/2 \leq t \leq 1$.
It should be pointed out that correlations lead to stronger localization off the band center, while they lead to
 critical delocalization at the band center.

We hope that our work initiates more studies on the ensemble of
correlated random band matrices with a general behavior of $B(N)\sim
N^s$, $C(N)\sim N^t$ where 
 $s,t$ may vary between $0$ and $1$, and  to reach  solid statements about the localization 
behavior in the thermodynamic limit.
We also like to point out, that the ''standard quantum Hall case'' of the correlated random band matrix models
 is not only a simple matrix
realization for quantum Hall systems, but has a very interesting distinction from other representative  
models for the quantum Hall universality class (for an overview over such models see \cite{Zirn99}). The correlated random band matrix
does not incorporate any handedness related to   the magnetic field. This handedness is essential in 
all other representative models that allow for the existence of extended states.
 In the correlated random band matrix model, however, the connection 
to a quantum Hall system goes via the Landau representation, which takes the handedness into account. 
Fortunately, the question of localization and delocalization is not restricted to that representation. 
In the correlated band matrix model 
the correlation of matrix elements is  the key for the quantum Hall transition. It would be very interesting 
 to construct a manageable field theoretic formulation for the correlated random band matrix model. 
This may be possible when taking advantage of the fact that the correlations are given by constraints
which may  be included by Lagrangian multipliers.

{\bf Acknowledgment.} MJ thanks B. Shapiro for previous collaboration on matrix models related to the quantum Hall effect
and  for  the central idea to 
construct the ensembles of  correlated random band matrices. We thank  J. Hajdu,
B. Huckestein, F. Izrailev and I. Varga  for useful
discussions.
 This research was supported in part by the
  Sonderforschungsbereich 341 of the DFG  and by the MINERVA
foundation.




\begin{thebibliography}{99}
\bibitem{Note}
This is in contrast to a quantum Hall strip of width $L$ and much larger length $L_y$, for which
the potential correlation length $d< \lambda$. In that case
one knows from many numerical simulations that the quasi-1D localization length $\xi_{\rm Q1D} (L)$
is of the order of  $L$ for critical states (see e.g. \cite{H94}).
\end{thebibliography}
\end{document}